# Combining Twitter and Mobile Phone Data to Observe Border-Rush:

# The Turkish-European Border Opening

Carlos Arcila Calderón[1], Bilgeçağ Aydoğdu[2], Tuba Bircan[3], Bünyamin Gündüz[4], Onur Önes[4], Ali Albert Salah[2], Alina Sîrbu[5]

[1] University of Salamanca, [2] Utrecht University, [3] Vrije Universiteit Brussel, [4] Turkcell Technology, [5] University of Pisa

The authors are listed alphabetically as they contributed equally to the work.

**Corresponding Author:** Tuba Bircan, tuba.bircan@vub.be

**Abstract**

Following Turkey's 2020 decision to revoke border controls, many individuals journeyed towards the Greek, Bulgarian, and Turkish borders. However, the lack of verifiable statistics on irregular migration and discrepancies between media reports and actual migration patterns require further exploration. The objective of this study is to bridge this knowledge gap by harnessing novel data sources, specifically mobile phone and Twitter data, to construct estimators of cross-border mobility and to cultivate a qualitative comprehension of the unfolding events. By employing a migration diplomacy framework, we analyse emergent mobility patterns at the border. Our findings demonstrate the potential of mobile phone data for quantitative metrics and Twitter data for qualitative understanding. We underscore the ethical implications of leveraging Big Data, particularly considering the vulnerability of the population under study. This underscores the imperative for exhaustive research into the socio-political facets of human mobility, with the aim of discerning the potentialities, limitations, and risks inherent in these data sources and their integration. This scholarly endeavour contributes to a more nuanced understanding of migration dynamics and paves the way for the formulation of regulations that preclude misuse and oppressive surveillance, thereby ensuring a more accurate representation of migration realities.

## Introduction

Following the death of 33 Turkish troops in airstrikes in Northern Syria on February 27, 2020, the Turkish government announced that refugees would no longer be stopped at the Turkish borders to Europe, and the doors are open[1]. This was immediately followed by movements of thousands of people towards the regional hub of Edirne in north-eastern Turkey, from where both the Bulgarian and Greek border crossings are easily accessible. In the following days, several declarations have been made by political leaders and media, stating that over 100,000 refugees left the border town of Edirne[2], although the International Organisation of Migration (IOM) and the United Nations High Commissioner for Refugees (UNHCR) estimated the number of people gathered at the border crossing points to be around 13,000[3]. Despite these and other official declarations[4], there is no clear picture of the movement patterns across borders in that period.

This research is based on the current theoretical and methodological discussion of how human mobility can be measured with digital traces in the era of big data (Fiorio et al. 2021, Tjaden 2021).

---

[1] https://www.bloomberg.com/news/articles/2020-02-27/erdogan-chairs-emergency-security-meeting-amid-syria-clashes
[2] https://www.indyturk.com/node/148101/haber/edirne-valili%C4%9Finden-s%C4%B1n%C4%B1r%C4%B1-ge%C3%A7en-s%C4%B1%C4%9F%C4%B1nmac%C4%B1-a%C3%A7%C4%B1klamas%C4%B1-son-iki-haftada-yakla%C5%9F%C4%B1k-150
[3] https://www.iom.int/news/more-13000-migrants-reported-along-turkish-greek-border
[4] https://www.theguardian.com/world/2020/feb/29/erdogan-says-border-will-stay-open-as-greece-tries-to-repel-influx

On one hand, Migration modelling is being challenged by the diversity of inputs generated from current migration processes, with individuals leaving more digital traces than ever. This new landscape may transform these traces into an explanatory variable of human migration. On the other hand, ss the inclusion of digital traces becomes standard in migration measurement, it's crucial to test these new data sources in different contexts, including prompted border crossings.

In this paper, we aim to assess new data sources, namely Twitter and mobile phone data, to measure mobility at the borders by focusing on the case of Turkish-European border opening in March 2020. Accordingly, we investigate a multi-modal approach to answer several questions of interest related to the border-rush in March 2020. The research questions, specific to the case at hand, are:

1. Which migrant groups attempted to cross the borders towards Europe following the opening of the border controls in Turkey?
2. What are the mobility patterns of these people in the time period of the border-rush?

In this paper, we also ask a series of methodological questions:
3. Can aggregated and anonymised Twitter and mobile phone data be used to gain insights into the border-rush?
4. What are the possibilities and limitations of these modalities, including the ethical and privacy related concerns?

We propose to detect mobility patterns via mobile extended detail records (xDR), and cross-check these with Twitter to validate and interpret our findings. Our analysis provides insights into the moving groups, their origin, and their behaviour both during and after the border-rush.

**Leveraging Issue-Linkage in Migration Diplomacy as a Capability**

Turkey, as the main transit country of the Eastern Mediterranean migratory route to Europe, has a significant role in European strategies for the management of migratory flows. On 18 March 2016, following the massive Syrian influx to Europe started in 2015, the Council of Europe and Turkey reached an official agreement on a legally non-binding statement, namely EU-Turkey Statement and Action Plan[5], which is most widely known as the *EU-Turkey Deal*: Syrian refugees to be resettled amongst EU member states and a promise of six billion euros financial support for Syrian refugees in Turkey between 2016 and 2019. This deal brought further negotiations on reopening Turkey's EU accession chapters and visa liberalisation (for Turkish citizens) dialogue. This joint collaboration is one of the most recent and clear examples of issue-linkage diplomacy as the entire process intently involves a bargaining mechanism "…where issues like asylum policies are linked to other issues, like trade, military and development policies" (Lemberg-Pedersen, 2016: 143).

While the EU stated its aim as improving the humanitarian situation, the deal obliged Turkey to keep the borders closed to prevent more refugees from entering Europe. While the EU leaned towards funding Turkish border control to prevent migration flows through the Eastern

---

[5] https://www.consilium.europa.eu/      4

Mediterranean Route by using issue-linkage strategies as a diplomatic tool, Turkey engaged in both cooperative and coercive migration diplomacy as a tool for bargaining. Both Turkey and the EU stayed committed to the deal, although the arrival of the financial support was delayed to 2018 and some negotiation points were not realised (Batalla Adam, 2017). Particularly, Turkey's accession negotiations and easing of the visa processes for Turkish citizens were stalled, due to EU's concerns about Erdoğan's government turning increasingly authoritarian. On February 11, 2020, weeks before the death of 33 Turkish troops in airstrikes in Northern Syria, the Turkish president publicly stated that "We do not have the word idiot written on our foreheads. We will be patient, but we will do what we have to do"[6]. On February 27, 2020, above-mentioned Turkish mechanised infantry battalion became the target of an airstrike in southern Idlib, and the Turkish government explicitly displayed its discontent about the progress of the EU deal. The public announcement on migrants to be allowed to pass through without border controls, mobilised thousands of people to reach the Turkey-EU border.

Considering the lack of reliable statistics on irregular migration, predicting the duration that migrants and refugees spent in the transit countries such as Turkey is not known, yet aspirations-capabilities framework argues that people are only likely to migrate when they can turn their aspirations into capabilities (Carling and Schewel, 2018; De Haas, 2021). We argue that the EU-Turkey issue linkage bargaining became an intermediary for aspiring migrants and refugees who leveraged the uncontrolled gates and developed an organisational capability to gather information on possible options for crossing the Turkey borders to Europe.

---

[6] https://www.bbc.com/

Due to the technical challenges of estimation, the number of people gathered and crossing the borders following Erdoğan's statements in February 2020 is still a matter of speculation. Therefore, investigating potential sources and methodologies to improve the estimations for acute mobility patterns. Accordingly, in this paper, we conceptualise the link between the migration diplomacy and issue-linkage nexus (Betts, 2010) and the aspirations-capabilities framework (De Haas, 2021) to understand the rush at the Turkish-Greek border following the political declarations in late February about revoked border controls at the European borders of Turkey, against the EU-Turkey deal.

**Measuring Mobility**

Migration as an overall demographic phenomenon is particularly difficult to measure (Poulain and Perrin 2008), and numerous long-lasting shortcomings have been identified (Ahmad-Yar and Bircan, 2021; Bircan et al., 2020; Laczko, 2020). Registered data collected at the international borders reflect the documented (regular) migration and fail to cover the undocumented (irregular) migration, which can happen at any point/part of the border line besides the official crossing points. Information and data about irregular migration are considered not only highly uncertain, but also insufficient and disputed. Despite the solid interest at the scientific, societal, and political levels, this knowledge gap presents a significant challenge for addressing the quantitative dimension of cross-border mobility[7].

---

[7] The Bureau of the Conference of European Statisticians (https://unece.org/) defines cross-border mobility as the continuous and temporary movement of people within a territorial demarcation between countries, which implies the crossing of a border

Considering cases such as the border-rush, nowcasting is not feasible through traditional data collection methods, even though the situation requires an immediate response that can only be facilitated through better estimation of the cross-border mobility. There are evident theoretical and methodological discussions on measuring human mobility through digital traces in the era of big data (Salah et al., 2022; Tjaden, 2021). On one hand, migration modelling and theorisation is being challenged by the diversity of inputs generated from the current migration processes, in which the persons on the move leave more (digital) traces than ever in history and sometimes can even narrate their crossings in real time. This is thus a new landscape for migration theories, given that these traces may become themselves an explanatory variable of human migration (i.e., when the traces boost the traditional contagious effect). On the other hand, even when there is not a consensus of validity of new data sources to measure mobility, the inclusion of digital traces will become a standard in migration measurement. Hence, there is a need to test these new data sources in different contexts, not only in steady migration routes but also in prompted border crossings. Consequently, unconventional data sources, such as social media data and mobile phone data can be utilised to enhance migration measures (for details: Salah et al., 2022), particularly for challenging cases like the border-rush at the Turkish-Greek border in March 2020.

*Migrant Mobility and Mobile Phone Data*

Mobile phone data can provide very detailed summaries of human mobility within a single country, and several earlier initiatives illustrated ways of ethical and privacy-aware processing of such data to gain insights without jeopardising private information about individuals (Blumenstock, 2012; Salah et al., 2019). In this work, we operate within a "data collaborative" with a telecommunications operator (i.e., Turkcell), and use data that are primarily collected by the operator for accounting and marketing purposes. Two types of data—call detail records (CDR) and mobile extended detail records (xDR)—inform our analyses. CDRs detail each subscriber's calls and messages, including time, source and destination IDs, base stations for location, and call duration. xDRs log data exchanges and 'handshakes' between phones and base stations. Content isn't examined; the focus lies on base station location, approximating the phone's location within Turkey's 780K km$^2$ area spread across around 4K base stations. This leads to a rough area of 200 km$^2$ per phone location, denser in cities and sparser in rural areas.

The xDR can be processed in an anonymised and aggregated fashion to enable a detailed analysis of human mobility. This is typically performed within a single country, as such data typically comes from a single telecommunication operator, and there is no record of base stations outside the country. Furthermore, because of high roaming charges, people moving across borders often stop using their old phone lines.

The use of mobile phone data to understand the mobility of migrant groups is a growing field of interest both in academia and among policy makers. The mobile phone data is highly useful for migration research, but it is challenging to obtain access rights. Existing studies are based on the data collected either through data collaboratives, or via private data sharing agreements with the telecommunication companies. Mobile phone data has characteristics that enable researchers to develop fine grained indicators of mobility, segregation, and integration (based on certain demographic characteristics such as gender, and nationality), identifying migrants, and mapping populations (Salah et al., 2019). The data can be stored in different formats depending on the means and purpose of data collection process.

Some earlier studies used mobile phone data for inferring patterns of internal migration, via computed indicators such as home and work locations, radius of gyration, the maximum distance travelled, or the number of cell towers used for groups of users (Blumenstock 2012; Hong et al. 2019). Since the behaviour of people are correlated with several indicators of interest, such as wealth or employment type, these measures can provide insights on internal migration. A special case of using CDR data for understanding refugee mobility was the Data for Refugees (D4R) Challenge, which made a dataset created from one million users available to the research community (Salah et al., 2019). The studies using this dataset compared the mobility behaviour and the call patterns of refugees and natives to provide insights and discuss solutions on pressing problems of refugees in Turkey. The D4R Challenge, and several challenges that preceded it, examined issues of privacy in detail, and proposed ways of processing data anonymously and in an aggregated fashion to protect the data subjects.

The most important disadvantage of mobile phone data is the difficulty of accessing it. A comprehensive legal framework needs to be established, and ethics committee approvals need to be obtained for each specific usage. We discuss the main points in the ethics section. Furthermore, there are certain biases that need to be factored in. In case the data come from a single telecommunication company, the total coverage is a fraction of the population (which is slightly over 40% for our study). As children do not legally own a phone line, they are not properly represented in the mobile data, which also needs to be considered.

*Human Mobility and Twitter Analysis*

The second source of information we tap in this paper is social media. Compared to traditional methods of inquiry (such as surveys), social media offer a dynamic and open space to study human mobility and migration, in a non-invasive way. Of all the social media available, Twitter stands out for being an important open tool that allows data collection for scholars and practitioners interested in studying both migrant mobility and the attitudes of local citizens towards newcomers and is our choice of platform for this study.

Locating a person through a Twitter message (i.e., Tweet) is possible if the person is using Twitter on a mobile phone, and if the geotagging option is enabled. Such messages can be used for detecting migrant stocks and trajectories. One relevant example is the work by Hawelka et al. (2014), who used geolocalised Twitter messages to identify global mobility patterns and to estimate the number of international travellers based on their place of residence. Another example is the study conducted by Zagheni et al. (2014), who used geolocalised Twitter messages to predict the inflection points of migration and to have a better understanding about migratory movements in the OECD countries. More recently, Kim et al. (2020) showed that general migration stocks can be extracted from Twitter data, by inferring a residence and nationality for Twitter users. They subsequently studied the social network for migrants for evaluating social integration (Kim et al. 2022a, b).

However, only very few messages are geotagged and contain enough metadata to estimate the location where the Tweet was produced (about 1% to 3% of the total stream, as stated by Morstatter and Liu (2017)). This is a major limitation when studying mobility via Twitter. Furthermore, Twitter users are a fraction of people on the move, and Twitter has a lower penetration rate compared to mobile phones and other social media platforms. This paper investigates whether Twitter data can provide insight into the mobility patterns at the border during events such as border-rushes.

While locations and mobility are difficult to extract from Twitter, a large body of scientific literature uses Twitter data to explore the sentiments and public representation of migrants and refugees (Coletto et al., 2016). Sentiment analysis via natural language processing (NLP) tools is used to analyse the tone of Tweets referring to newcomers and migrants, or to detect racism (Chaudhry, 2015) or even hate speech (Müller and Schwarz, 2018) against them. However, there are certain biases that need to be considered. Criss et al. (2020) argue that Twitter is a social medium in which many people may feel emboldened to publish offensive messages based on the perception of anonymity, and frequently, it serves as an "echo-chamber," where similar points of view (even if they are racist or xenophobic) are amplified.

Various studies investigated the negative representation of migrants in social media, and Twitter in particular (Gallego et al., 2017; Kreis, 2017). In this sense, one of the recurring topics in this type of speech is an apparent need to defend the society against this "threat," which is represented through hashtags such as #StopInvasion or #refugeesnotwelcome (Kreis, 2017). Messages about migrants and refugees that circulate on Twitter may identify them as a burden and/or threat to the receiving society (Díaz and Gualda, 2017; Amores and Arcila, 2019). On the other hand, there will also be messages of support and solidarity.

In this paper, we also perform sentiment analysis of hashtag data from Twitter to understand the overall sentiment and its changes during the border-rush.

**Data and Methodology**

Alternative data sources and new methodologies have been used by numerous scholars to complement the shortcomings in migration statistics (Robinson and Dilkina, 2018; Alexander et al., 2020). Yet, considering the border mobility, the examples are scarce (for exceptions: Massinen et al. 2019; Silm et al. 2021). Our study particularly assesses mobile phone data and Twitter data to estimate the mobility patterns at the Turkish-European border in March 2020.

*Mobile Phone Data*

The mobile phone data are collected by Turkcell, in the format of xDR, which are used as billing records of the subscribers. These records include transactions, sessions, and call information, and are registered with a timestamp and information about the service providing base station, which translates to a rough location of the subscriber. We processed the xDR collected by Turkcell in Turkey during 2020, and demographic data collected during the subscription of the subscribers. Based on the nationality of subscribers, we created two groups: *Visa* group consisting of foreigners (who need a visa to enter Bulgaria and Greece), and *No-Visa* group (those who do not need a visa). For the purposes of our study, we filtered the foreign subscribers who have been at the Western border provinces of Turkey, Kırklareli and Edirne, at least once between the 25th of February and the 25th of March 2020. We filtered top 20 foreign nationalities that have the largest number of subscribers in our sample, and as a result around 38,000 foreign subscribers were selected.

As our purpose with this data set is to gain insights on the mobility of selected subscribers, we fetched their xDR data between February 28 and June 15. xDR is highly valuable data source in case of border rush as it shows mobility of subscribers in real time. Spatially, we aggregated the data at the province level, whereas temporally, aggregation was done at a daily level (i.e., 24-hour blocks). As subscribers can appear in more than one province per day, each user is assigned to the province in which they get the most signals for a given day. In case a subscriber does not receive signals for a few days, we assume that they stay in the district from where they received their last signal. Any subscriber, who does not appear in the data from a specific date till the end of the period of observation, is labelled as a "lost subscriber". Before the aggregation, these subscribers are dropped from the data set after the last day observation. In addition, we also used antenna level data to observe the general patterns of signals at the border provinces. The antenna level data does not contain any individual level information. It is simply showing the count of devices that are receiving signals from the antenna in real-time.

We are specifically interested in people who leave the country and have no further transactions after a point. Our approach in processing the mobile phone data have a few issues with respect to giving accurate estimates on the number of foreigners who might have left Turkey during this period. The "lost subscriber" label is a noisy indicator, and it is only suggestive in terms of its implications. Firstly, the numbers are likely to be exaggerated simply because the subscribers that we assume who left the country might have churned to another mobile operator, or they might have been left unused within the period of tracking. Secondly, the daily aggregation is making it difficult to observe short term trips of subscribers to border provinces. We might be falsely assuming some individuals left the country from the border provinces, whereas they were just on short term trip to the border provinces. Therefore, we neither claim that "lost subscribers" were actually lost, nor we claim that they left the country, we simply label them as being disappeared from the mobility data set at certain date earlier than June 15, 2020.

*Twitter Data*

We have collected Twitter data from the 25th of February to the 25th of March 2020 (the same initial period of xDR data above), using the Academic Track Twitter API. Specifically, we searched for geolocated tweets at the European Turkish border, using the set of bounding boxes shown in Figure 1. From the set of tweets obtained we selected only the those posted on Turkish territory. This resulted in a total of 10,684 tweets, posted by 1,460 users. For these users, we also downloaded all the tweets up to the end of 2020. We henceforth denote this dataset as the *geolocalised Twitter dataset*.

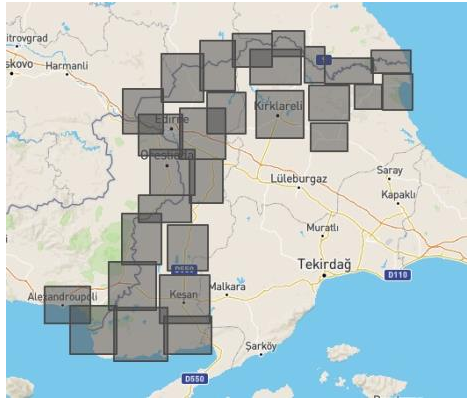

**Figure 1:** Bounding boxes used for data collection for the geolocalised Twitter dataset.

The analysis of the geolocalised Twitter dataset is divided into two parts. We first study the initial period between February 25 and March 25, by inspecting the languages employed by the users present at the border. Most of the tweets downloaded already contain language information, added by the Twitter platform itself. Some tweets, however, are not assigned a language: the API returns the "und" language label, standing for "undefined". Closer inspection shows that along with a lot of http links, a majority of these tweets include fragments and hashtags that are Turkish words or names, often with some characters replaced with other symbols. For these tweets, we assign a language based on the majority language of the user. Specifically, for each user we count the tweets in each language, and all their undefined tweets are assigned the language with most tweets. In this way we label almost completely the tweets with an undefined language. Only 58 tweets remain unlabelled in the period February 25 - March 25 and these are removed from the analysis. With the resulting data we compute the aggregated number of tweets and users for three different language groups: *Visa*, *No-Visa*, *Turkish*. The Visa/No-Visa groups consist of all languages present in our data that are spoken in countries where visas are/are not required to enter the EU.

The second analysis concentrates on the destination of the users and employs tweets posted from May to December 2020. We combine destinations into three classes: *Europe* including the European Union, United Kingdom and other Balkan and Eastern European countries, *Turkey*, and the rest of the world as *Other*. We divide the tweets by destination, and then for each destination we further divide them into subgroups based on language groups as above. We are interested in the number of users present in each of the subgroups obtained, i.e., how many users visited each destination and tweeted in each language group. The undefined tweets are labelled as before, leaving only 8 tweets unlabelled (omitted from the rest of the analysis).

In addition, during a similar period (February 28 to the March 31) we collected tweets from the same Twitter API filtering by specific hashtags, manually curated, that we deemed highly representative of the situation. We retrieved 158,002 messages in English (37,195), Greek (2,317), Arabic (290) and Turkish (118,200) including the selected hashtags[8]. These hashtags were chosen because they were trending topics at the beginning of the crisis or because they were highly representative of the situation. We denote this dataset as the *hashtag dataset*.

**Findings**

*Mobility Patterns from Mobile CDR*

Following Turkish officials' announcement on February 28th about opening borders, events escalated at the four border gates between Edirne and Greece/Bulgaria, and one gate from

---

[8] #IStandWithGreece, #Yunanistan, #suriye, #suriyeli, #multeci, #refugees, #refugeecrisis, #syrianrefugees, #RefugeesWelcome, #göçmensorunu, #avrupabirliği, #HumanRightsRefugee, #suriyelileriistemiyoruz, #negülüyorsunerdoğan, #SenGülkiÜlkenGülsünReis, #Greekborder, #GreeceAttacksRefugees, #GreeceUnderAttack2, #sınırKapıları, #ipsala, #turkishborder, #kapılaracıldı.

Kırklareli to Bulgaria. Daily reports from Greek and Turkish media reveal the Evros River as a common crossing site, with a significant rush to the Pazarkule gate near Edirne. Refugees were reportedly amassed in the dead-zone between Greek and Turkish borders, though official data on crossings and migrant composition remains unavailable.

In our analysis, we divided foreign subscribers into two categories: Visa and No-Visa groups. The Visa group includes prominent nationalities such as Syrians, Afghans, and Iraqis who require a visa to enter Greece and Bulgaria, while the No-Visa group comprises Greek, Bulgarian, and Moldovan nationals who don't require a visa.

Figure 2 shows the temporal variation of these two groups based on their location, either at the border or in other cities, along with a category for lost subscribers. A noticeable surge in Visa-required subscribers at border cities is observed on February 28 and 29. The figure also marks three dates, namely March 10, April 4, and May 14, where there was a sharp drop in subscriber numbers, mainly from border regions and especially Edirne. On May 14, subscriber numbers declined both from border cities and Istanbul.

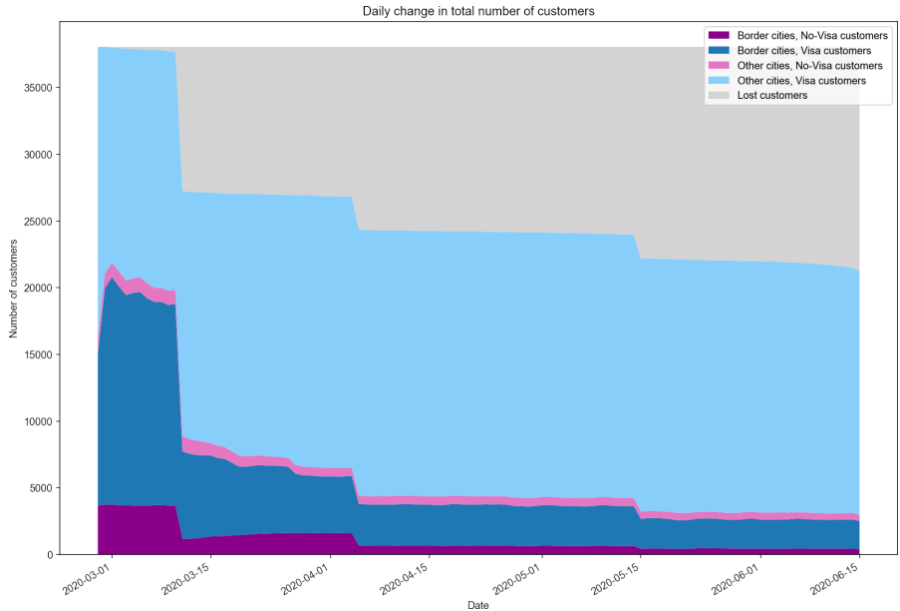

**Figure 2:** Temporal variation of selected customers in five group; the combination of Visa group, and No-Visa group either at the border provinces or other cities, and lost customers.

In Figure 3, we can see the presence at the border cities in the beginning of the rush, and on June 15. The total number of subscribers was still high in June at the centre of Edirne. This can be due to the fact that a proportion of selected subscribers were already living in that province. Another observation we can make is that the border-rush was towards Edirne and Greek borders, rather than Bulgarian borders. The provinces at the Northern-Eastern provinces in the map had a very similar number of subscribers during and after the border-rush, whereas the number of subscribers at the Greek borders, were much larger. These observations are in line with the qualitative evidence we could find from various news sources of the date, which focused on the river of Evros as the main line of border crossings.

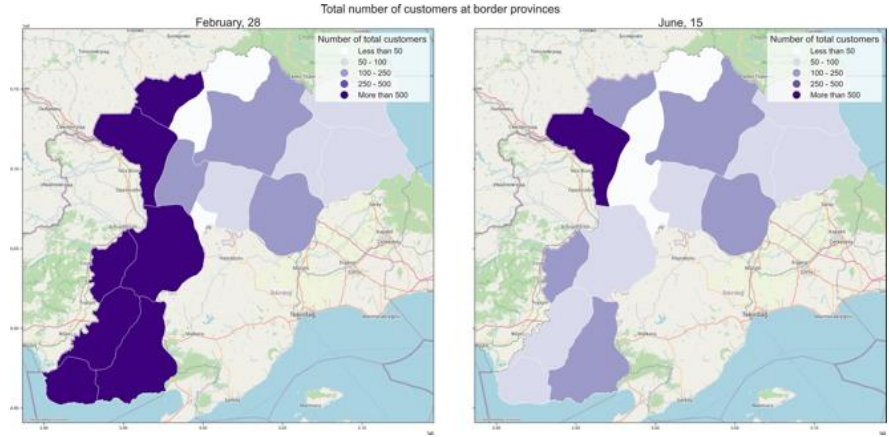

**Figure 3:** Total number of customers at the border provinces, on February 28, and June 15.

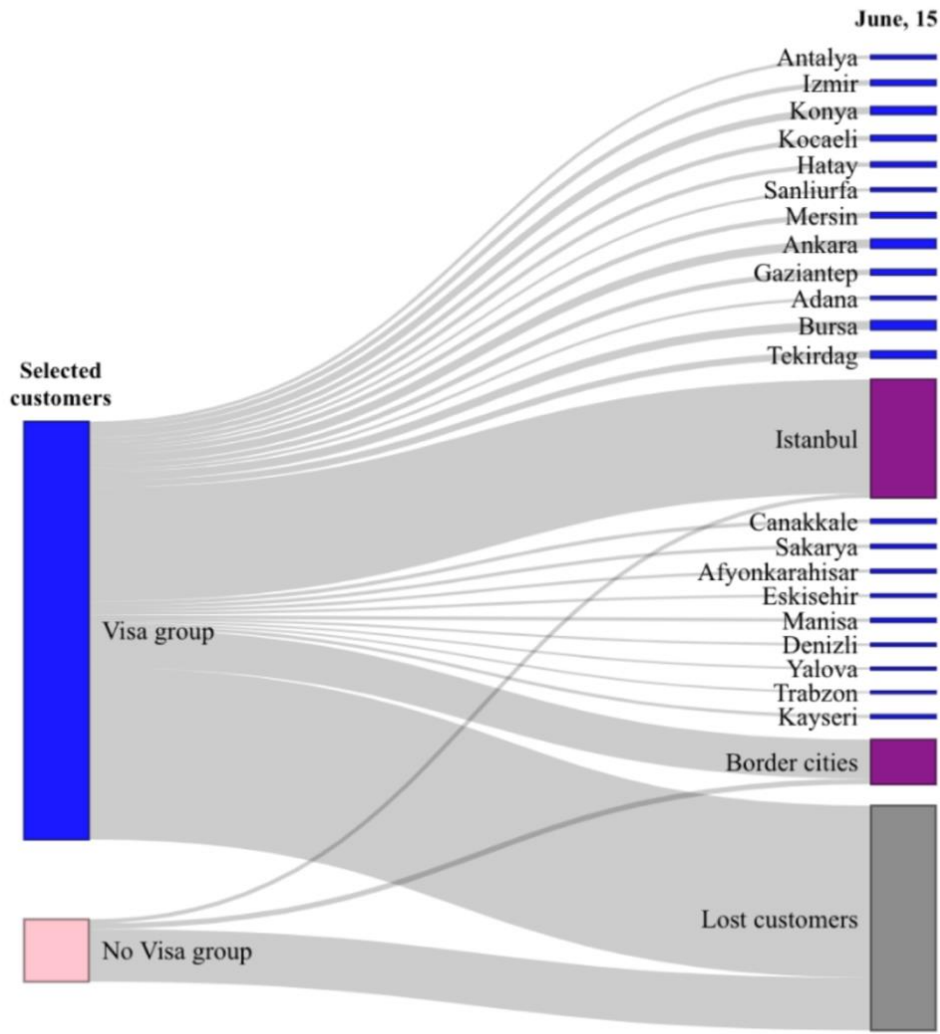

**Figure 4**: Sankey diagram shows the destination city of selected customers.

In Figure 4, a Sankey diagram describes the change in location between February and June of different subscriber groups from the border to other internal Turkish regions. It shows that almost half of the selected subscribers were lost during the period. Among the No-Visa subscribers, who were still actively using their mobile phones in June, they were all either in the city centre of Edirne, or in the capital district of Istanbul, Fatih. For the Visa subscribers though, we see great variety in terms of their destination provinces. Many remained in the border cities or were in Istanbul, and others were spread around the country.

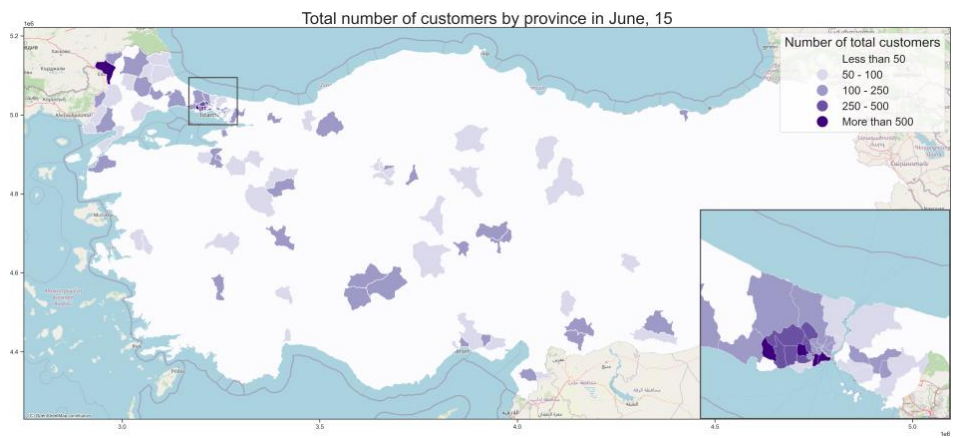

**Figure 5:** Total number of customers by province on June, 15. The figure only shows the distribution of people who were at the European border during the border rush in March, but stayed in the dataset after the rush. The city of Istanbul is zoomed for better visibility in the lower right corner.

Note: The figure only shows the distribution of people who were at the European border during the border-rush in March but stayed in the dataset after the rush. The city of Istanbul is zoomed for better visibility in the lower right corner.

Figure 5 indicates the bulk of subscribers moved from Istanbul to border provinces on June 15, with Konya, Gaziantep, and Hatay being other popular destinations. We can see that the distance to the border provinces played a role in terms of the decision to go to the border during this period;

most subscribers travelled to the border provinces from Istanbul. During the peak of the border-rush, approximately 3,000 No-Visa and 17,000 Visa group subscribers were at the borders, a noticeable increase from around 3,000 and 6,000 respectively on January 1.

During the border-rush peak, around 3,000 No-Visa and 17,000 Visa group subscribers were at the borders. On January 1, the border cities hosted around 3,000 No-Visa and 6,000 Visa subscribers. The selected population of the No-Visa group is likely to be the residents of these cities, whereas a large proportion of the Visa group has come to the border cities in January and February, especially before the rush. The 'lost subscriber' label, indicating no signal from the phone, denotes a small percentage of subscribers disappearing daily, but large losses from border provinces suggest significant border crossings post-March 2020. Both subscriber groups showed similar patterns, with March 10 having the most disappearances. Based on our qualitative research (on Twitter conversations below) we could not find any specific event that might have caused a disproportionate amount of border crossings on that day.

These figures, however, don't provide a precise migrant count. Our data only includes Turkcell subscribers, representing a 40% telecom market share (with 32,71 million subscribers) in Turkey, but an estimated 50% among main migrant groups (Aydogdu et al., 2021). Irregular migrants, potentially underrepresented in the data, may have participated more in the border rush. On the other hand, our methodology, labelling any subscriber disappearing from data as having left the country, could overestimate departures. Due to these factors, we can't provide an exact population count but estimate that 10,000-20,000 Visa and 2,000-4,000 No-Visa subscribers crossed the borders in March.

*Border Presence and Sentiment from Twitter Data*

Our first analysis of the geolocalised Twitter dataset combines the space, time, and language dimensions. We study the languages at the European borders, to investigate whether Twitter data can provide insights into the different groups present there.

From a language perspective, the population present at the border in the period of February 25 to March 25 is very heterogeneous, with 35 languages present. Figure 6 summarises the number of tweets in the different language groups: Visa, No-Visa, Turkish. We note a very strong presence of Turkish tweets, which is understandable given that we are looking at the Turkish territory. No-Visa group follows, while the Visa group is the smallest, with only 206 tweets over the observed period. This may indicate that the Visa group is the smallest, or that the members of this group do not make use of geolocalised tweets. If we compare this with the results from the mobile data discussed above, we observe that there the Visa group is larger than the No-Visa group. One explanation for this could be that the patterns we see in the Twitter data are due to low usage of geolocalised tweets by users in the Visa group. Also, the users in the No-Visa group could be underrepresented in mobile xDR data if they continue to use their foreign mobile operators in Turkey.

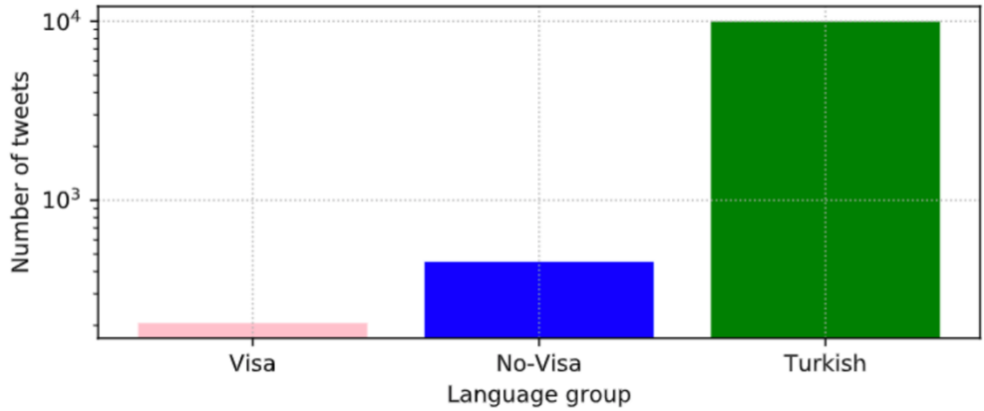

**Figure 6:** Number of tweets posted between the 25th of February and the 25th of March 2020 in different language groups at the European border.

Furthermore, we study the number of individual users tweeting in each language group, as shown in Figure 7. The Turkish language maintains the top position, while for the Visa group, we obtain only 97 users in total.

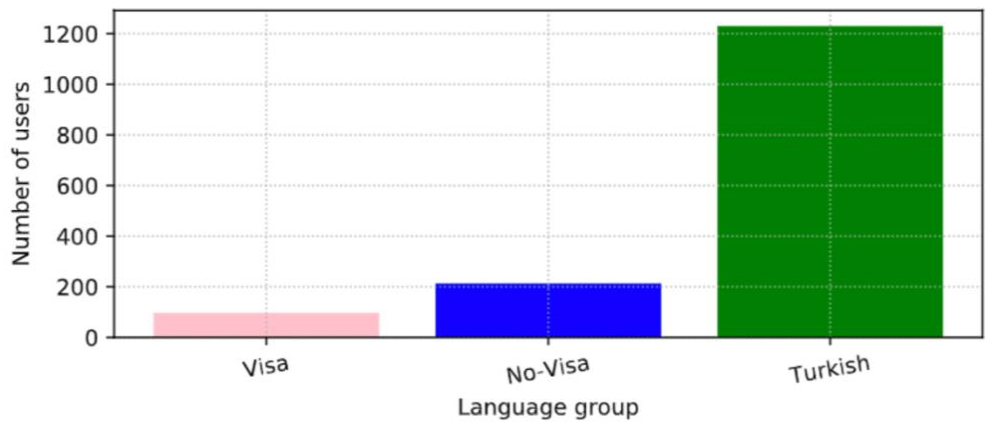

**Figure 7:** Number of users posting between the 25th of February and the 25th of March 2020 in different language groups at the European border.

Additionally, we analyse the temporal patterns for the three language groups. Figure 8 shows the number of daily tweets at the EU border. A first observation is that the daily numbers are quite reduced. The Turkish tweets always dominate the discussion. While generally the trends are stable, we can observe some local peaks in March: on the $2^{nd}$, $5^{th}$, $9^{th}$, $11^{th}$ for Visa group; on the $3^{rd}$-$6^{th}$ and $12^{th}$ for No-Visa group; on the $5^{th}$, $10^{th}$, and 12th for the Turkish group. These could be related to the peak in "lost" subscribers observed in xDR data on the March 10.

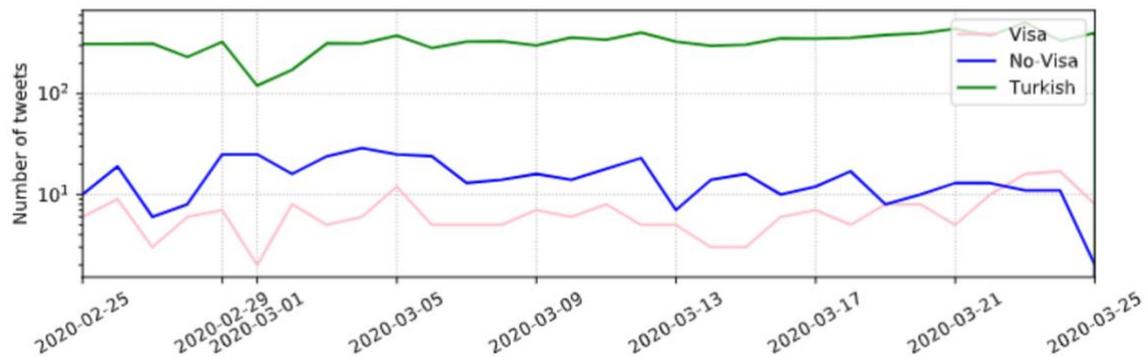

**Figure 8:** Daily number of geolocalised tweets at the EU-Turkish border, for different language groups.

We next try to understand the destinations of the users who were seen at the border during the initial period of analysis in March. We consider all tweets posted by the 1,460 users in the months May-December 2020 and extract only the language group and location of these tweets. We note that, in total, 943 out of 1,460 users have geolocated tweets in the period between May and December, while the others disappear from the data. For each language group and location, we count the number of individual users.

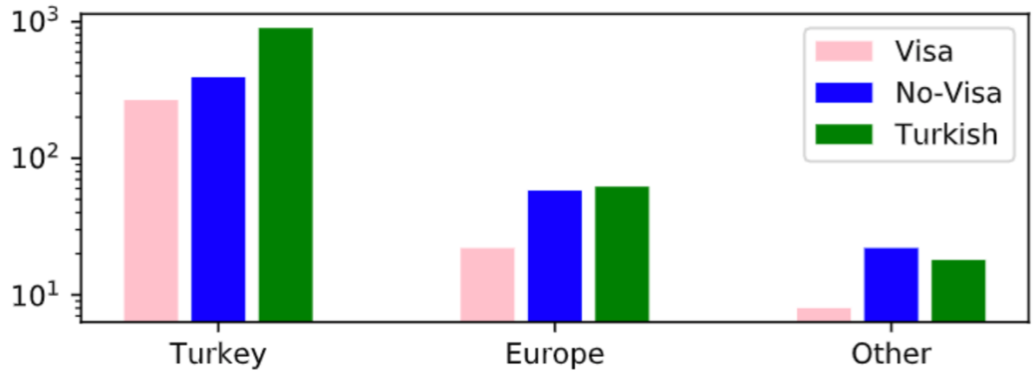

**Figure 9:** Destinations for all four groups geolocated between May and December.

Figure 9 shows the size of the various groups obtained. We note that most users were located in Turkey during the period observed, and this applies for all language groups. The second destination is Europe, again for all language groups, with comparable absolute values. However, proportionally, we may say that a larger fraction of Visa users is present in Europe.

We underline the fact that in this analysis we simply study the overlap between language and destination, without tracing individual users. Therefore, if one user tweets in more languages, and travels in more destinations, they are counted in each of them. To evaluate the extent of the double counting, we show in Figure 10 the overlap between the various subsets of users, obtained by splitting the 943 users by language group and destination. We note that only about a half of the users use a single language only, and it applies mostly to Turkish. Furthermore, there is a significant number of users who tweet in all language groups, making it difficult to use language as a proxy for country of origin. This could be explained by the fact that users could combine the use of Turkish (the language of their current country of residence), English, an internationally spoken language included in the No-Visa group, and their own language (Visa or No-Visa group). This is in fact something reported also by a recent qualitative study of migrants en route to Europe (Purkayastha and Bircan, 2023). In terms of location, the Venn diagram in Figure 10 shows again that Turkey dominates as a destination.

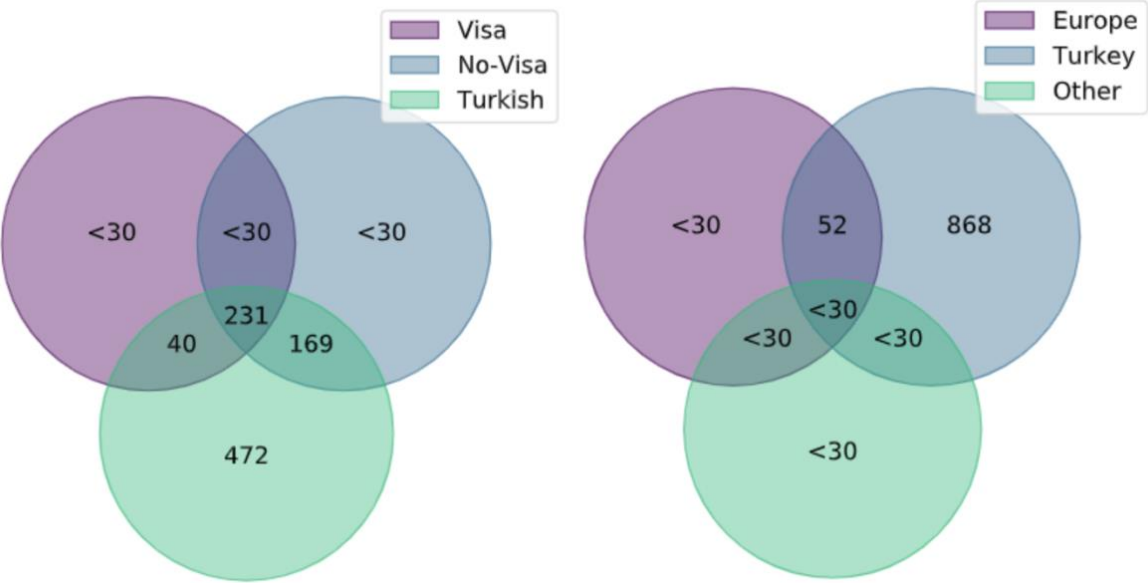

**Figure 10:** Overlaps in the set of users for different language groups and destinations (May-December).

All in all, these results suggest that, if any crossing of the border towards the Europe was performed, this was not only among the Visa group, but also among the Turkish and No-Visa groups. Importantly, we cannot make any assumptions about whether the crossings were regular or not, based on Twitter data. This is not surprising, as border crossings could also be due to work or leisure travel. Considering the socio-political and humanitarian aspects of the specific case we are investigating, in the area of interest, the presence of other actors (i.e., journalists, social workers and humanitarian aid volunteers) may also be expected, and these could be more likely to use multiple languages in their tweets.

Even if the Twitter data suggest that a larger fraction of the Visa language group (proportionally speaking) crossed the border, there is a lot of heterogeneity in terms of language and destination, in the sense that the same user employs multiple language groups and visits multiple destinations. Also, the absolute numbers are very small in terms of users, suggesting that maybe border crossings were not massive during that period. All in all, we can conclude that we cannot draw detailed quantitative information from such data, but only identify possible general trends.

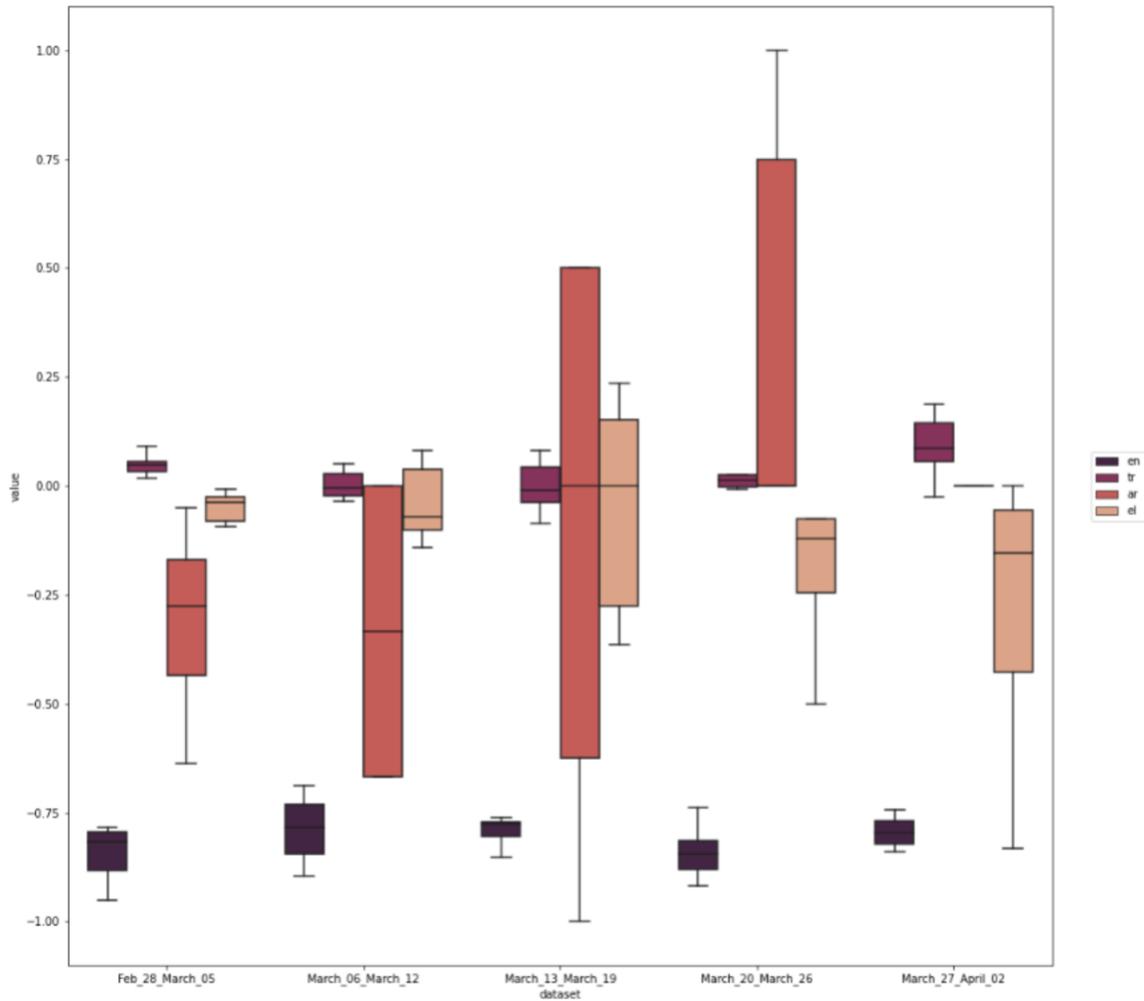

**Figure 11:** Sentiment distribution in each language, for each week.

Considering the sentiments, we utilised *SentiStrength* (Thelwall et al., 2010) to perform lexicon-based sentiment analysis on our hashtag dataset, a tool highly accurate in English and validated in other languages like Turkish (Vural et al., 2013). Each tweet received a sentiment score in a scale from -5 to +5, from which we computed daily and weekly averages. Weekly data showed stable sentiment levels across languages, excluding Arabic which displayed a significant negative peak during the second week and a positive peak in the fourth. Average sentiments were especially negative in English and Greek, neutral in Turkish, and varied in Arabic. This does not signify a prevailing negative view towards migration but possibly a general critique of various government policies. Our analysis of sentiment variance over time (Figure 11) suggests a polarised discussion in Arabic or Greek, whereas English and Turkish sentiment variability was low.

Extreme sentiment words were scarce in our tweets, with English (0.0147% of 543,792 words) and Turkish (0.0082% of 1,750,278 words) featuring them, but none in Greek or Arabic, indicating that strong sentiment words in Arabic or Greek did not influence the observed variability. As our hashtag dataset was globally sourced, the sentiments reflect varying international and national opinions on the border-rush. English and Greek users showed a mostly negative conversation about the event, while Turkish and Arabic users displayed more neutral or sometimes positive conversations, revealing greater sentiment polarisation in Greek and Arabic.

*Twitter Conversations and Loss of Mobile Phone Subscribers*

To analyse the correlation between Twitter conversations and spikes in lost mobile subscribers (March 10, April 4, May 14), we conducted content analysis on tweets from the geolocalised and hashtag datasets. We compared the discussions during these peaks to the remainder of the period.

We examined tweets by 1,326 users present at the borders on the aforementioned dates, finding two types of users: (1) a minority of people on the move and (2) a majority of Turkish citizens living at the border.

Migrants mostly sought updates about border conditions, shared data with other migrants, their families, or friends, and expressed their experiences as migrants. They used words related to meeting points, border status, jails, roads, cities, and changes in the migration surveillance systems.

The second group, the Turkish citizens living at the border, primarily discussed political matters, frequently converging on migration views. Some demonstrated religious convictions, referencing right-wing leaders, and exhibiting negative sentiments towards migration, often using derogatory terms (e.g., "invaders"). Others were more government-critical, articulating concerns about migration-related impacts on economy, agriculture, commerce, and workforce. Both groups exhibited clear negativity towards the EU's handling of migration, leading to an overall negative discourse on migration during these periods.

Then, we examined the We scrutinized tweets from a hashtag dataset during three prominent spikes: March 9/10, April 3/4, and May 14/15. These primarily Turkish tweets (95%-100%) were categorised into topics to examine their association with media events.

The first spike coincided with President Erdoğan's visit to Brussels (March 9) regarding the migration crisis. Greek (4.7%) and Turkish (95.3%) tweets were analysed separately. 56% of Greek tweets cited the European border opening, seen in polarized hashtags like #GreeceUnderAttack (50%) and #WirhabenPlatz (57%, "we have space"). On March 10, Greek tweets mainly supported their government via hashtags such as #GreeceUnderAttack (56%) and #IStandWithGreece (56%). Turkish tweets focused on Edirne's border crisis (3.3%-2.8%), and "Başkanımız" ("Our President") was a key topic on March 10 (2.3%).

The second spike, not associated with a specific media event, saw Turkish messages focus on border openings (4.7%) with terms like "Edirne" (2%) and "Kırklareli" (2.7%-3%). Tweets on April 4 also mentioned the pandemic and impending lockdown with "evden" (2.8%, "from home"). The third spike followed public condemnations of Turkey's "illegal acts" on its borders by France, Greece, Egypt, Cyprus, and the UAE (2020). Turkish tweets (99% of sample) on May 14 mostly mentioned border cities Kırklareli (3%) and Edirne (2.3%). On May 15, #AkşenerDemokrasiArenasında (2.8%) trended, referring to opposition politician Meral Akşener's TV appearance.

Our analysis revealed that during days with notable subscriber losses, the three potential user groups (migrants, border citizens, or the public) focused on mobility issues, mostly negatively portraying migrants. Even when official statistics and media showed no special mobility patterns, Mobile CDR and Twitter data identified particular events augmenting human mobility.

**Ethical, Legal, and Privacy Issues**

The people we study in this paper are vulnerable. Subsequently, while we aim to understand mobility patterns to inform policy decisions, we adhere to several ethical guidelines to ensure no harm results from data processing, and there is no surveillance or processing of personal information.

We followed the principle of "privacy by design and default." Accordingly, we anonymised and aggregated the data to a point where it was impossible to identify or track individual users before we began the analysis. For mobile phone data, this involved spatio-temporal aggregation, and for home locations, we aggregated at the city level.(De Montjoye et al. 2013). De Montjoye et al. 2013). Access to raw data and analysis were separated by a "Question and Answer" model, where raw data remained on the servers of the telecommunications operator, and only aggregated indicators were available for analysis.

For Twitter data, we aggregated in time, space, and language, so users could not be re-identified from the published results. The qualitative analysis was performed anonymously, considering only the tweet text and no information about the users or locations.

The measures we took to protect the privacy of users are not restricted to the results published in this paper, but also cover the intermediate steps of the analysis. Considering mobile data, raw data were never accessed by our researchers. In the case of Twitter, public raw data were downloaded using the public API and stored securely. These data went through a process of anonymisation and minimisation as a first processing step, before studying mobility. So, for each tweet we extracted, only those fields that were necessary were employed in the analysis: the location (at country level), the period (initial March period or May-December), the language, and the pseudonymised user IDs. This allows protecting the identity of users also during the analysis, and reduces the possibility of de-anonymising, especially given the fact that the same raw Twitter data can be in principle downloaded by any other Twitter developer with an academic license. In the case of the qualitative analysis, minimisation and anonymisation were achieved by employing only the tweet text and the day of posting.

All these measures are intended to protect not only the privacy of individuals, but also group privacy, which is another important aspect. Except for Turkey, which is the main country of the events and where we have a larger number of users, no other individual country is identified. We always group languages and locations, for both Twitter and Mobile data, so that different groups of migrants cannot be identified and therefore cannot be profiled and targeted based on our analysis. Furthermore, the Mobile and Twitter IDs are not matched and combined in any way.

Finally, we have opted to wait for a significant amount of time to publish the results of the analysis. Especially with mobile data, rapid processing may have implications for the vulnerable populations (such as indication of illegal crossing points); we follow the "do no harm" principle and add the time buffer for an extra measure, and we are careful in our analysis to employ a level of spatial aggregation to prevent such a use.

**Discussion**

The Syrian civil war, subsequent humanitarian crisis, and the Covid-19 pandemic are significant 21st-century events. The EU and Turkey pursued strategic issue-linkage negotiations, leading to short-term concessions for Turkey and migration management strategies for the EU. Turkey's dissatisfaction with the EU's delayed, selective responses led to repeated threats to rescind the agreement.

The March 2020 Turkey-Greece border rush, a societal reaction to discourse on easing border controls, dominated European and Turkish narratives. Estimations of those who left Turkey varied significantly based on data sources. Assessing societal reaction to political change during these crises is challenging due to data scarcity and the swift, irregular mobility of individuals trying not to miss the "open gates".

In this paper, we demonstrated the strengths and weaknesses of alternative data sources for assessing the sudden human mobility at the European borders of Turkey, as well as how they can complement each other to cover various angles of related complex societal issues. Our findings show that the numbers of people crossing into Europe from the border circulated in the mainstream media[9] may be inflated, and the demographic distribution may be different than what is shared in these channels. Specifically, mobile data indicate that the actual numbers are smaller (i.e., closer to the number of people estimated by IOM). Furthermore, the analysis of both mobile and Twitter data indicate that refugees are not the only groups who crossed the border during the border-rush. Moreover, we illustrate that mobile data can give good quantitative indicators of mobility and has good coverage, but it says very little about the reasons of mobility and needs to be combined with other data sources. Twitter, on the other hand, has very little coverage (especially when only the geo-tagged Tweets are considered), and is not suitable for estimating the amount of mobility during border-rush, but contains qualitative insights as the event unfolds.

---

[9] https://www.sozcu.com.tr/

The number of persons who crossed the borders could have been influenced by the COVID19 pandemic co-occurring with the border opening. However, in Turkey the first COVID case was registered on the 11$^{th}$ of March, so after the day that our data indicates as the major border crossing moment. Furthermore, restrictions in Turkey were generally lighter than most of the rest of the world, with night curfews, school closing and public transportation limitations, and most stringent in the month of April, so after the border rush. Therefore, we expect that the effect on the rush was reduced.

As elaborated above, this novel approach has the advantage that it can provide timely insights on trends and qualitative information from Twitter data. Besides these purpose-driven approaches for the Big Data analytics, a few limitations, and trade-offs of the data-driven approaches of the methodology should be recognised to ensure the correct interpretation of the findings. The first limitation regards the sampling bias in social media data (Hargittai 2020). Particularly, the results of the geolocated Twitter data analysis covered few users -given the specificity of the event-, therefore, although the findings indicate observed trends, they do not provide sufficient evidence for quantitative analysis and estimate development in this case. One potential explanation can be the lower Twitter penetration in areas where refugees come from[10]. Additionally, it is possible that some Twitter users do not enable geolocation, and that could be especially true for some category of users who may not want to disclose their location. Another limitation is in the interaction between user behaviour and the assumption we make. While some mobile phone or Twitter users may be very active, so they appear in the data very often and their location is accurate, some other users may use their phones rarely, making them disappear from the data for long time periods. This may interfere with our assumption that a user who has disappeared from the data has left the country, inflating thus our estimates. The methodology we presented benefits a lot from the presence of active users but may suffer in the presence of less active ones.

---

[10] https://www.statista.com/

Moreover, an important disadvantage is that unlike other applications on managed migration estimates such as migrant stocks and flows (Sîrbu et al., 2021), there is no gold standard for rapid cross-border mobility, so a proper validation of the results is very difficult. Given the time-lag in the availability of the migration statistics (Ahmad-Yar and Bircan, 2021), one would need to wait for a few years for official data to appear, to validate the conclusions drawn. Even then, the available official statistics might not give the absolute numbers, since irregular migration is one of the most difficult types of international migration to record (Bircan et al., 2020). One can only try to make corrections based on user shares and provide estimates as accurate as possible, as we did with mobile data.

The methodology and data types presented were aimed at studying border rush, however they can also be employed to study migration in general. Part of the customers who disappear from the data for long time periods can be assumed to have moved abroad, however one needs to distinguish those who have changed mobile phone operator. Seasonal migration could be also observed by studying the alternation between active and non-active periods. For Twitter, some studies of migration and mobility already exist (Sîrbu et al, 2021), however with the same advantages and disadvantages that we discussed above. Twitter data can be useful for those countries with a large user base; however, they are weak in countries where Twitter is not used.

The last consideration is the possibility of governments and international organisations harnessing the big data analysis methodologies for ever-more restrictive border controls. While we took many precautions (and ethics approval) for our own study, governments can create a legal basis to process sensitive data for their own purposes. Only proper research on socio-political dimensions of human mobility into possibilities, limitations, and risks of these data sources (and of their combination) will allow regulations to be created to prevent misuse, and oppressive surveillance.

**Data Availability Statement**

The datasets generated for this study are available publicly. All data used are anonymised and aggregated to ensure the privacy and confidentiality of individual data contributors. These datasets can be accessed at the following link:

https://github.com/bircantuba/HumMingBird-Project.git

DOI: 10.5281/zenodo.11211701

**Acknowledgements**

This work has been supported by the European Commission through the H2020 European project "HumMingBird – Enhanced migration measures from a multidimensional perspective" (GA: 870661), and partially by the H2020 project "SoBigData++: European Integrated Infrastructure for Social Mining and Big Data Analytics" (GA: 871042).